\documentclass{osa-article}

\journal{oe}

\articletype{Research Article}

\usepackage{xr}
\usepackage{soul}
\usepackage{amsmath}
\usepackage{graphicx}
\usepackage{epstopdf}
\usepackage{subfig}
\usepackage{url}
\usepackage{float}
\usepackage{bm}
\usepackage{ifthen}
\usepackage{braket}

\def\Pr{\mathrm{Pr}}
\def\sD{{\bm{\mathcal{D}}}}
\def\bz{\mathbf{z}}
\def\imag{\mathrm{i}}
\DeclareMathOperator{\Tr}{Tr}
\def\sZ{\mathcal{Z}}

\usepackage[absolute]{textpos}

\usepackage{xcolor}

\begin{document}
\title{Bayesian homodyne and heterodyne tomography}

\author{Joseph C. Chapman,\authormark{1,$\dagger$,*} Joseph M. Lukens,\authormark{1,$\dagger$} Bing Qi,\authormark{1,2} Raphael C. Pooser,\authormark{1} and Nicholas A. Peters\authormark{1}}

\address{\authormark{1}Quantum Information Science Section, Oak Ridge National Laboratory, Oak Ridge, TN 37831\\
\authormark{2}Now with Cisco Systems, Inc., San Jose, CA 95134\\
\authormark{$\dagger$}contributed equally}
\email{\authormark{*}chapmanjc@ornl.gov}

\begin{abstract*}
Continuous-variable (CV) photonic states are of increasing interest in quantum information science, bolstered by features such as deterministic resource state generation and error correction via bosonic codes. Data-efficient characterization methods will prove critical in the fine-tuning and maturation of such CV quantum technology. Although Bayesian inference offers appealing properties---including uncertainty quantification and optimality in mean-squared error---Bayesian methods have yet to be demonstrated for the tomography of arbitrary CV states.
Here we introduce a complete Bayesian quantum state tomography workflow capable of inferring generic CV states measured by homodyne or heterodyne detection, with no assumption of Gaussianity. As examples, we demonstrate our approach on 
experimental coherent, thermal, and cat state data,
obtaining excellent agreement between our Bayesian estimates and theoretical predictions.  
Our approach lays the groundwork for Bayesian estimation of highly complex CV quantum states in emerging quantum photonic  platforms, such as quantum communications networks and sensors.

\end{abstract*}


\section{Introduction}

\label{sec:intro}

The types of photonic states typically encountered in quantum information science (QIS)---including quantum networking, computing, and sensing---can be divided into two broad categories, depending on the fundamental unit of information employed. In qubit (or more generally, qu$d$it) encoding, logical states are defined by single photons which occupy $d$ electromagnetic modes; in this case, photonic QIS protocols are typically defined for a specific numbers of particles~\cite{Knill2001, Kok2007}. In qumode encoding, however, the electromagnetic modes themselves assume fundamental priority, with information encoded in, e.g., field variables such as the quadratures or the number of photons~\cite{Braunstein2005, Weedbrook2012, Pfister2020}. Now, because qubit-encoded optical states are typically available in a probabilistic fashion only---due to spontaneous generation processes~\cite{Shih2003}, post-selection by design~\cite{Knill2001, Kok2007}, or simply from loss---the qumode, or continuous-variable (CV), formalism is arguably the more general of the two viewpoints, for it accounts for variable numbers of photons and subsumes discrete-variable (DV) qubit/qudit encoding as a special case.

This generality yields a significant increase in complexity. For example, whereas a single qubit is fully described within a two-dimensional Hilbert space, an infinite-dimensional Hilbert space is required to specify a single qumode. Although this dimensionality can be truncated in practice with reasonable assumptions about maximum photon number, fully general quantum state tomography (QST) of a qumode~\cite{Lvovsky2009} remains a considerably more computationally intensive affair than QST of a qubit~\cite{James2001}. Based on a collection of homodyne or heterodyne measurements with a local oscillator, numerical techniques for qumode QST have progressed from linear inversion with the Radon transform~\cite{Smithey1993a, Smithey1993b} to maximum likelihood estimation (MLE)~\cite{Banaszek1999, Lvovsky2004}, compressive sensing~\cite{smith_quantum_2013,kyrillidis_provable_2018}, and, very recently, neural networks~\cite{Tiunov2020}.

\begin{figure*}[tb!]
\centerline{\includegraphics[width=5.25in]{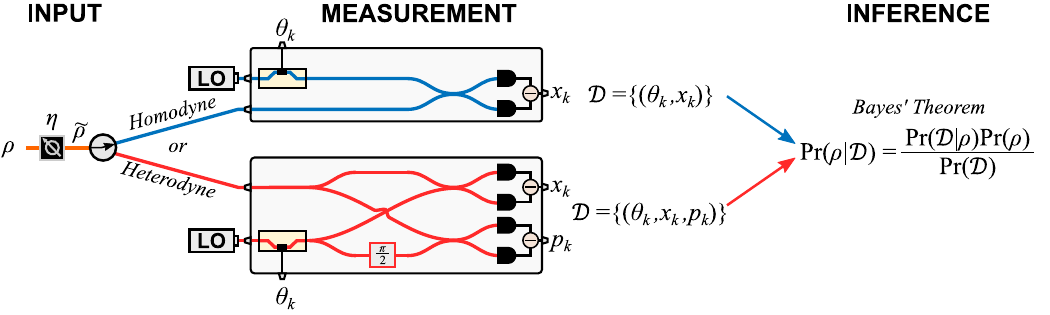}}
\caption{Bayesian QST workflow. An arbitrary single-mode quantum state with density matrix $\rho$ first undergoes loss, then is sent either to a homodyne setup (top), or to a heterodyne setup (bottom). Results $\sD$ are then fed into an inference algorithm based on Bayes' theorem, which returns the posterior distribution over possible density matrices, given the observations, $\Pr(\rho|\sD)$.} 
\label{fig:conceptHT}
\end{figure*}

In any method, one desires an estimate of the underlying quantum state which is as close as possible to the unknown ground truth. Bayesian techniques prove uniquely situated to reach this end~\cite{Robert1999, MacKay2003, Blume2010}. Given a generic dataset $\sD$ and to-be-determined quantum state $\rho$, Bayes' theorem expresses the possible states in terms of a probability distribution $\Pr(\rho|\sD) = \Pr(\sD|\rho)\Pr(\rho)/\Pr(\sD)$, which (i) provides automatic uncertainty quantification for any collection of measurements or outcomes and (ii) returns the optimal estimate of any quantity of interest: the Bayesian mean estimator formed as the mean over $\Pr(\rho|\sD)$ minimizes the squared error averaged over all states and outcomes~\cite{Robert1999}. Unfortunately, the high-dimensional integrals involved with full Bayesian inference only exacerbate the computational challenges of qumode QST. And while Markov chain Monte Carlo (MCMC) algorithms have facilitated Bayesian tomography of a variety of discrete qubit/qudit systems~\cite{Seah2015, Granade2016, Williams2017, Mai2017, Lukens2020b, Lukens2021a, Lohani2021, Lu2021} and continuous qumode states that are \emph{Gaussian}~\cite{DiGuglielmo2009}, no method for Bayesian inference of arbitrary CV states has been proposed.

In this work, we introduce and demonstrate a complete workflow for Bayesian QST of generic CV states. Capable of handling both homodyne and heterodyne measurements, our method combines physical models developed in previous CV estimation approaches with advanced MCMC algorithms~\cite{Cotter2013} for full Bayesian tomography with state-of-the-art computational efficiency. After detailing our method, we demonstrate inference on several experimental datasets, including newly measured coherent and thermal state examples and legacy cat state data from Ref.~\cite{Gerrits2010}. As the first Bayesian method for CV QST that does not assume Gaussianity in the model, our workflow closes the existing gap between CV and DV systems in Bayesian QST and provides new opportunities for informationally efficient tomography in state-of-the-art non-Gaussian photonic QIS.

\section{Model}
\label{sec:model}

Figure~\ref{fig:conceptHT} depicts the envisioned tomographic scenario. An arbitrary single-mode quantum state with density matrix $\rho$ first undergoes loss described by the efficiency parameter $\eta\in[0,1]$. Then it is either sent to a homodyne setup (top), where it is mixed with a local oscillator aligned to phase $\theta$ and measured with a balanced detector, or to a heterodyne setup (bottom), which divides the signal with a 50/50 beamsplitter and performs simultaneous homodyne detection of the outputs, where the local oscillator is aligned to $\theta$ or $\theta + \pi/2$. Because loss commutes with either homodyne or heterodyne detection~\cite{Qi:20}, non-unit detector efficiency can be absorbed into $\eta$ and the detection itself viewed as ideal; in the process, we do invoke the assumption that all detectors possess equal efficiencies---the desired design condition in practice.

While the ground truth state exists in an infinite-dimensional space, truncation is required for numerical tractability. To this end, we choose to express $\rho$ in the Fock basis $\ket{n}$, so that such truncation can be realized directly by imposing a photon-number cutoff $n_c$, i.e.,
\begin{equation}
\label{eq:rho}
\rho = \sum_{m=0}^{n_c} \sum_{n=0}^{n_c} \rho_{mn}\ket{m}\bra{n},
\end{equation}
corresponding to a total Hilbert space dimension of $D=n_c+1$. Writing the density matrix after loss as
\begin{equation}
\label{eq:rhoTilde}
\tilde{\rho} = \sum_{m=0}^{n_c} \sum_{n=0}^{n_c} \tilde{\rho}_{mn}\ket{m}\bra{n},
\end{equation}
the elements can be expressed in terms of a generalized Bernoulli transformation~\cite{Campos1989}
\begin{equation}
\label{eq:rhoTilde2}
\tilde{\rho}_{mn} = \sum_{k=0}^{\min(n_c-m, n_c-n)}B_{m+k,m}(\eta)B_{n+k,n}(\eta) \rho_{(m+k)(n+k)},
\end{equation}
with $B_{j+j',j}(\eta) = \sqrt{ {j+j' \choose j} \eta^j (1-\eta)^{j'} }$. At this point, we note that from a tomographic perspective, $\eta$ need not be bound to the actual system loss but can be viewed as an inference choice: $\eta$ represents whatever efficiency an experimenter is comfortable separating from the state itself. For example, $\eta=1$ can be selected as a conservative means to absorb all system inefficiencies into the state $\rho$, while $\eta<1$ will seek the state before the selected loss, accounting for the associated nonlinear transformation in Eq.~(\ref{eq:rhoTilde2}). (This freedom will be leveraged for the experimental comparison of homodyne and heterodyne tomography in Sec.~\ref{CHOHT}.)

This state is then mixed with a local oscillator (LO) with relative phase $\theta$ and detected. In the homodyne case, the quadrature measurements $x$ follow the probability density function~\cite{Lvovsky2009}
\begin{equation}
\label{eq:hom}
f_1(x|\theta,\rho) = \sum_{m=0}^{n_c}\sum_{n=0}^{n_c} \tilde{\rho}_{mn} \frac{e^{\imag (n-m)\theta}}{\sqrt{\pi m! n! 2^{m+n}}} e^{-x^2} H_m(x)H_n(x),
\end{equation}
where $H_n(x)$ is a Hermite polynomial and we follow the convention $\hbar=1$ so that the variance of vacuum is $\braket{\Delta x^2} = \frac{1}{2}$. In this formulation, we have not considered additional electronics noise; if appreciable in a given experiment, such noise can be absorbed into the transmissivity parameter $\eta$~\cite{Appel2007}. For heterodyne detection, a single measurement outputs a pair of quadrature values $(x,p)$ according to the density~\cite{Richter1998}
\begin{equation}
\label{eq:het}
f_2(x,p|\theta,\rho) = \sum_{m=0}^{n_c}\sum_{n=0}^{n_c} \tilde{\rho}_{mn} \frac{e^{\imag (n-m)\theta}}{\pi \sqrt{m! n!}} e^{-(x^2+p^2)} (x-\imag p)^m (x+\imag p)^n
\end{equation}
where $x$ is the output aligned to phase $\theta$ and $p$ to $\theta+\frac{\pi}{2}$. By expressing $\rho$ in the Fock basis and using general formulas for the probability density functions of the homodyne [Eq.~(\ref{eq:hom})] and heterodyne [Eq.~(\ref{eq:het})] measurements, our workflow is able to measure and reconstruct non-Gaussian inputs as well as Gaussian inputs.

In the tomographic context, these measurements are repeated $K$ times, leaving dataset $\sD =\{(\theta_k,x_k)\}$ (homodyne) or $\sD =\{(\theta_k,x_k,p_k)\}$ (heterodyne) where $k=1,2,...,K$. Each phase value $\theta_k$ is allowed to vary but is assumed known from independent calibration. The likelihood $L_\sD(\rho) \propto \Pr(\sD|\rho)$ follows as the product of individual outcomes:
\begin{equation}
\label{eq:LLhom}
L_\sD(\rho) = \prod_{k=1}^K f_1(x_k|\theta_k,\rho)
\end{equation}
for homodyne and
\begin{equation}
\label{eq:LLhet}
L_\sD(\rho) = \prod_{k=1}^K f_2(x_k,p_k|\theta_k,\rho)
\end{equation}
for heterodyne. While beyond the scope of this investigation, we note that any mixture of homodyne and heterodyne measurements---analogous to ``time sharing'' in the theory of Gaussian channel capacities~\cite{Takeoka2014}---can be handled as well, through a likelihood consisting of a combination of $f_1$ and $f_2$ factors.

Up to this point, no aspects of this model are unique to Bayesian inference, but instead are common to any estimation procedure built on the likelihood. Indeed, MLE proceeds precisely by finding the physical density matrix $\rho$ which maximizes $L_\sD(\rho)$ above [Eq.~(\ref{eq:LLhom}) or Eq.~(\ref{eq:LLhet})]. In the Bayesian paradigm, however, the likelihood is accompanied by a prior distribution as well, often chosen to be as uniform as possible to give reasonable weight to all possible states in the Hilbert space. Examples of fiducial density matrix distributions, studied extensively in DV encoding, include Hilbert--Schmidt and Bures, the former equal to the distribution of $D\times D$ density matrices found by tracing out random pure $D^2$-dimensional states~\cite{Zyczkowski2001}, and the latter noteworthy as the only monotone distribution that is both Fisher and Fubini--Study adjusted~\cite{Sommers2003}.

To our knowledge, no comparable analyses of density matrix distributions exist for the CV case, leaving wide open the question of fiducial prior selection. Since the photon cutoff $n_c$ [cf. Eq.~(\ref{eq:rho})] should exceed the maximum photon number of all states of interest, it certainly would seem reasonable to consider a prior which preferentially weights lower photon numbers over higher ones---in contrast to uniform DV priors that place all logical basis states on equal footing. Nonetheless, given the absence of specific alternatives in the literature, we elect to implement a Bures prior distribution in the truncated $(D=n_c+1)$-dimensional space for our inference examples here. In applying uniform weight to all photon numbers up to $n_c$, Bures can be viewed as a highly cautious distribution that will permit low-uncertainty estimates only when justified by the data collected.

To construct density matrices drawn from the Bures distribution, we first define a $D^2$-dimensional vector of complex parameters $\bz=(z_1,...,z_{2D^2})$; the first $D^2$ populate a $D\times D$ matrix $G$, while the second $D^2$ parameters are used to form another complex matrix that is converted to a $D\times D$ unitary $U$ through the Mezzadri algorithm~\cite{Mezzadri2007}. The density matrix constructed according to
\begin{equation}
\label{eq:Bures}
\rho = \frac{(I_D+U)GG^\dagger (I_D+U^\dagger)}{\Tr\left[(I_D+U)GG^\dagger (I_D+U^\dagger)\right]},
\end{equation}
where $I_D$ is the $D\times D$ identity, will observe a Bures distribution if each parameter $z_j$ is independently drawn from a complex standard normal distribution
~\cite{Osipov2010}. While an ``overparameterization'' in that each density matrix is expressed in terms of $4D^2$ independent real numbers---compared to the minimum of $D^2-1$ required by physicality~\cite{James2001}---this construction is the most efficient known for the Bures distribution, and has enabled demonstrations of Bayesian inference with Bures priors up to $D=64$ in DV examples~\cite{Lu2021}.

The posterior density can therefore be written as
\begin{equation}
\label{eq:posterior}
\pi(\bz) = \frac{1}{\sZ} L_{\sD}(\bz) \pi_0(\bz),
\end{equation}
a convenient form of Bayes' theorem emphasizing the functional dependencies on the parameters $\bz$ only. Here $\pi_0(\bz) \propto \prod_{j=1}^{2D^2} e^{-|z_j|^2/2}$, $\sZ$ is a normalization constant such that $\int d\bz\,\pi(\bz)=1$, and the equivalence $L_\sD(\bz)\equiv L_\sD(\rho(\bz))$ is understood. With a collection of $R$ samples $\{\bz^{(r)}\}$ drawn from $\pi(\bz)$ through Monte Carlo methods, the Bayesian mean estimator $f_B$ of any function $f(\rho)$ can be computed according to
\begin{equation}
\label{eq:BME}
f_B = \int d\bz\, \pi(\bz) f(\rho(\bz)) \approx \frac{1}{R}\sum_{r=1}^R f(\rho(\bz^{(r)})),
\end{equation}
circumventing the need for high-dimensional integration. The Bayesian mean density matrix---the point estimator chosen for density matrix and Wigner function plots below---is therefore defined as $\rho_B \approx \frac{1}{R}\sum_{r=1}^R \rho(\bz^{(r)})$.

Obtaining the samples $\{\bz^{(r)}\}$ represents the core computational bottleneck of Bayesian methods and has motivated decades of research in MCMC techniques~\cite{Robert1999, MacKay2003}. Recently, we applied a particularly efficient MCMC method, known ``preconditioned Crank--Nicolson'' (pCN)~\cite{Cotter2013}, to Bayesian QST~\cite{Lukens2020b}. Designed to eliminate the step size/acceptance rate tradeoff intrinsic to random walk MCMC, pCN was found in our tomographic examples to obtain significant speedups compared to alternative algorithms ~\cite{Lukens2020b}. This pCN-based Bayesian QST workflow has since been applied to several problems with a Bures prior distribution~\cite{Lu2021, Lohani2021}. We adopt the same sampling procedure here, with the only substantive difference being the likelihood computed: starting with a parameter vector $\bz^{(r)}$, we first compute $\rho$ according to Eq.~(\ref{eq:Bures}); then the lossy $\tilde{\rho}$ follows via Eq.~(\ref{eq:rhoTilde2}); and finally, the likelihood is evaluated according to Eqs.~(\ref{eq:hom}, \ref{eq:LLhom}) for the homodyne case, or Eqs.~(\ref{eq:het}, \ref{eq:LLhet}) for heterodyne.

\section{Experimental demonstration}
\label{sec:exp}
\subsection{Setup}
\begin{figure*}
\centerline{\includegraphics{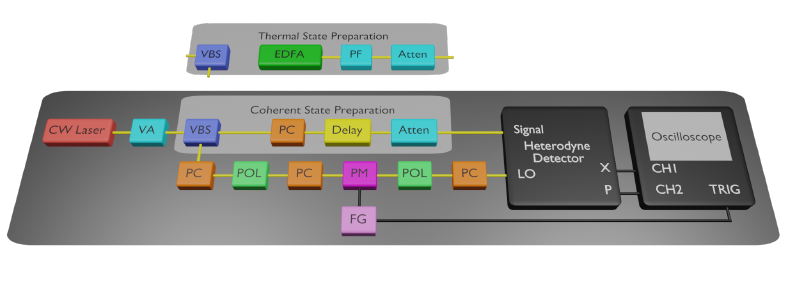}}
\caption{Heterodyne tomography setup for coherent and thermal states. For coherent states, the continuous-wave (CW) laser is split and used to make attenuated coherent states and a local oscillator (LO) for the tomography.  For thermal states, an erbium-doped fiber amplifier (EDFA) is filtered and attenuated to create thermal states of low average photon number. The CW laser is used as the LO in the tomography. Definitions: VA = variable attenuator, VBS = variable beamsplitter, PC = polarization controller, Atten = fixed attenuator(s), POL = polarizer, PM = phase modulator, FG = function generator, and PF = programmable filter.}\label{fig:expsetup}
\end{figure*}

Our setup for producing experimental tomographic datasets is summarized in Fig.~\ref{fig:expsetup}. To generate a local oscillator (LO), we use a continuous-wave (CW) laser at 1548.8~nm (Pure Photonics PPCL550), then send it through a phase modulator (EOSpace PM-0K5-10-PFA-PFA-DC), which is driven with a voltage ramp (8.3 Vpp, 5 kHz) from a function generator (Stanford Research Systems DS345) producing a full $2\pi$ phase swing, then to the LO input of the heterodyne detector (Optoplex RX-KC0100C821AC), which is a 90$^{\circ}$ optical hybrid with outputs connected to a pair of amplified balanced detectors. The two electrical outputs of the detector are 50~$\Omega$ terminated into a digital sampling oscilloscope (Keysight MSOX4104A) that we use to acquire a total of $K=7998$ measurement points $(\theta_k,x_k,p_k)$, where each quadrature value is the average of four consecutive 2.5~GHz samples on our oscilloscope, repeated at an interval of 100~ns,  for all examples.

To measure coherent test states (Fig.~\ref{fig:expsetup}), we split about 1\% of the CW laser power before the phase modulator using a variable fiber beamsplitter (Newport F-CPL-1550-N-FA) and send it through a polarization controller. It is subsequently passed through a variable optical delay line (General Photonics VDL-001-35-33-FC/APC-SS) and delay fiber to approximately match the path length to the LO, then through several fixed attenuators, and finally to the signal input of the heterodyne detector. Signal and LO path lengths are matched to within about 1~ns, which is much shorter than the approximately 0.1~ms coherence time of the laser. From the raw measurements, we calculate $g^2(0)\approx1$ using the analysis method in Ref.~\cite{Qi:20}, consistent with the production of high-quality coherent states.

To produce thermal states (inset in Fig.~\ref{fig:expsetup}), we use the amplified spontaneous emission from an erbium-doped fiber amplifier (Pritel SCG-40), filter it to 1548.8$\pm$0.4~nm (Finisar Waveshaper 1000A), and send it to the signal input of the heterodyne detector. The analysis method of Ref.~\cite{Qi:20} returns $g^2(0)\approx2$ for this case, as expected for thermal states. See Appendix A 
 for descriptions of our calibration and data processing procedures.
Finally, all Bayesian inference results below consist of $R=1024$ total samples, which were themselves obtained by thinning an MCMC chain of total length $RT$, where $T$ was selected empirically for convergence in the mean and standard deviation of the inferred fidelity.

\begin{figure}[bt!]
\centerline{\includegraphics[width=3.375in]{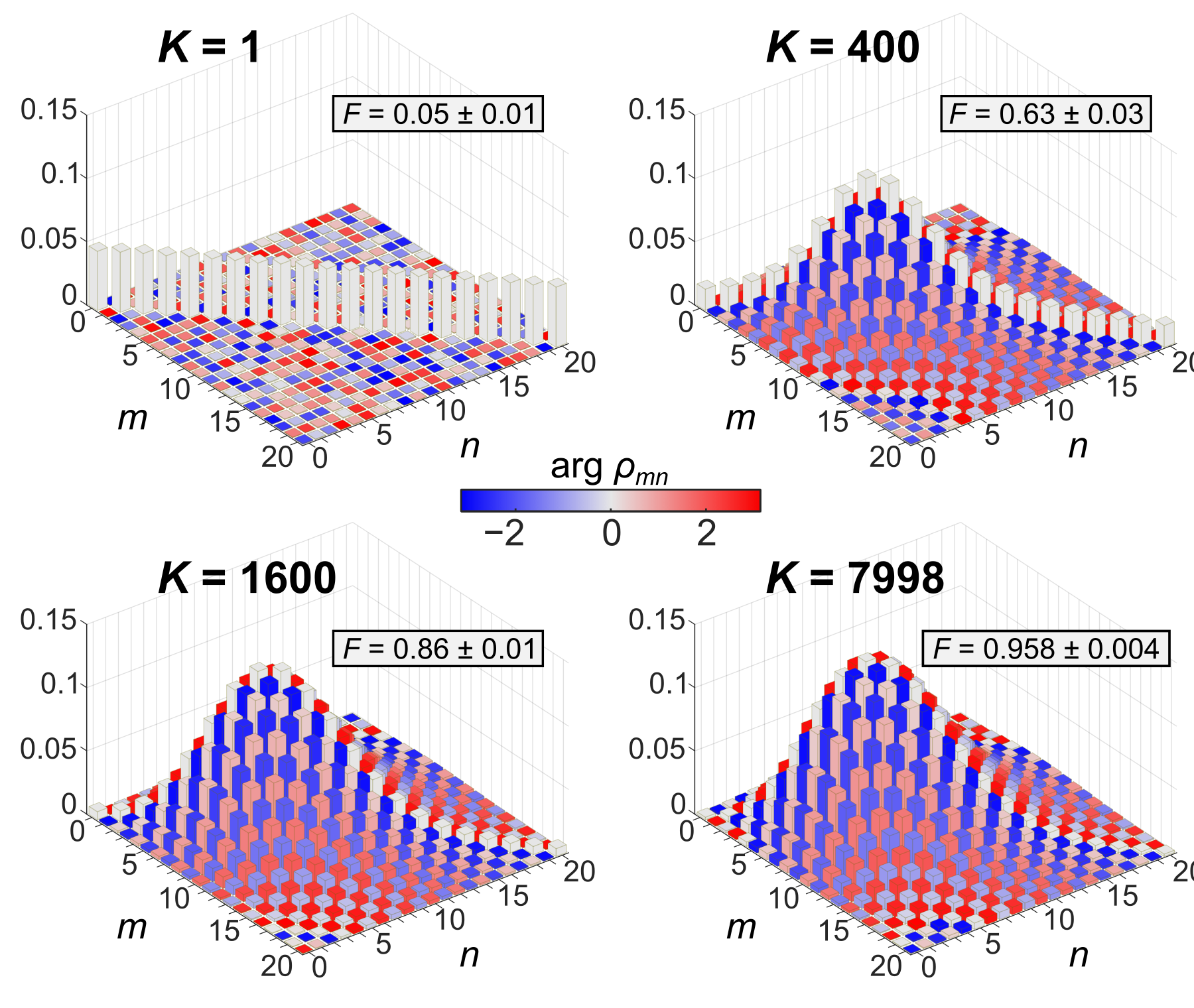}}
\caption{Scaling with the number of measurements for experimental heterodyne tomography. Retrieved Bayesian mean density matrices and fidelities for the case $|\alpha_0|^2=7.97$, for several values of the number of measurements $K$.}\label{fig:denmatprog}
\end{figure}

\subsection{Heterodyne inference}
We test our Bayesian method on six separate states of varying type and amplitude; the insets of Fig.~\ref{fig:denmats} plot the raw quadrature data in phase space $(x_k,p_k)$, with color signifying the respective LO phase $\theta_k$. The quadrature data of Fig.~\ref{fig:denmats}(a) show clear variation with LO phase, as expected for coherent states, while the data of Fig.~\ref{fig:denmats}(b) appear rotationally symmetric and remain centered at the origin, characteristic of thermal states. In order to provide a benchmark to compare against subsequent Bayesian results, we compute an expected coherent state parameter $\alpha_0$ or thermal state mean photon number $\mu_0$ by correcting for the LO phase and performing averages over the raw data: $\alpha_0=\braket{x_k \cos\theta_k - p_k\sin\theta_k} + \imag\braket{x_k\sin\theta_k + p_k\cos\theta_k}$ or $\mu_0 = \braket{x_k^2 + p_k^2} - 1$. These values are then used to define the expected ground truth states for fidelity calculations. In defining the states in this fashion, we are effectively taking $\eta=1$ in the model, absorbing any loss into the state itself, which has little impact in these cases since coherent and thermal states preserve their statistics under loss~\cite{Qi:20}. 

\begin{figure*}[tb!]
\centerline{\includegraphics[width=3.375in]{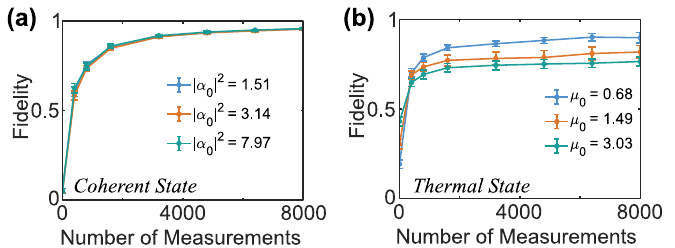}}
\caption{Fidelity scaling with heterodyne detection. (a)~Fidelity of inferred states with respect to expected ground truth for an attenuated laser (coherent state). (b)~Inference fidelity for amplified spontaneous emission (thermal state).}\label{fig:COHvsTHERM}
\end{figure*}

We took these data and produced inferred density matrices with our Bayesian QST workflow, using an LO power of 12 mW, an MCMC thinning value of $T=2^{14}$, and taking our photon cutoff at $n_c=20$;  for the expected states examined, this selection ensures a truncation error $\epsilon = 1-\sum_{n=0}^{n_c} \braket{n|\rho_g|n}$ of at most 0.003 (for the $\mu_0=3.03$ thermal state); the truncation error for all coherent states is less than $10^{-4}$. In Fig.~\ref{fig:denmatprog}, we show an example of how the elements of the Bayesian-mean estimate $\rho_B$  progress as more measurements $K$ are included in the likelihood, for the attenuated laser dataset with $\alpha_0 = -2.78 -0.54\imag$ ($|\alpha_0|^2=7.97$). At a single measurement $K=1$, virtually no information has been gained about the ground truth state, and $\rho_B$ averages to the maximally mixed state in the 21-dimensional space, as needed for the Bures prior. For $K=400$, the result clearly resembles a coherent state, with a residual diagonal spike from the prior, and by $K=1600$ the estimate matches the expected ground truth with fidelity $F=0.86\pm0.01$, improving to a fidelity of $F=0.958\pm0.004$ at $K=7998$, which we expect would continue to increase with even more measurements.

\begin{figure*}[tb!]
\centerline{\includegraphics[width=5.25in]{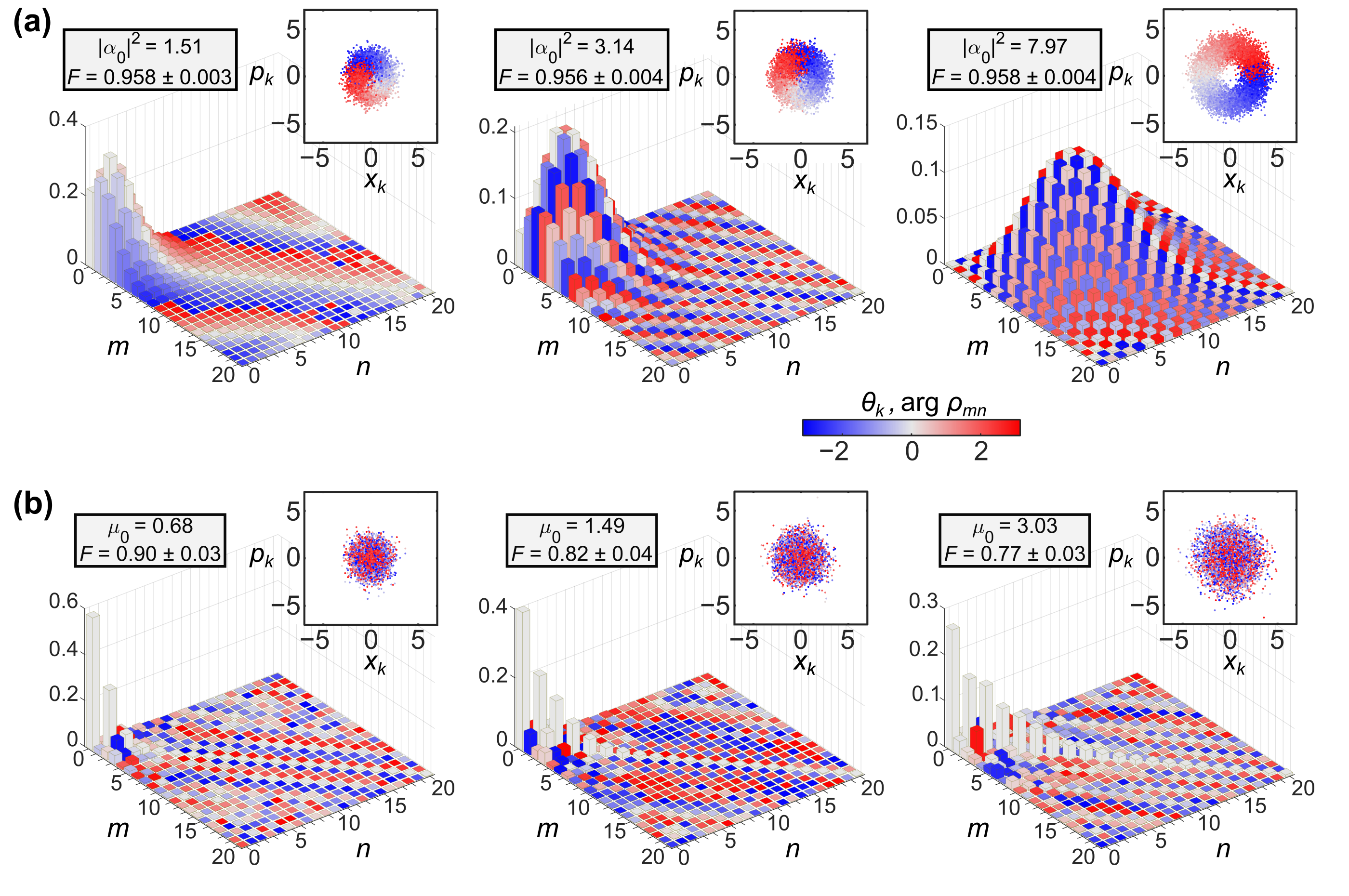}}
\caption{Bayesian QST with heterodyne detection.  (a)~Bayesian mean density matrices estimated from all 7998 measurements of each attenuated laser dataset.  (b)~Bayesian mean density matrices using all 7998 measurement results for each tested thermal state. Insets show the raw data points $(x_k,p_k)$ in measurement phase space, with color indicating the LO phase $\theta_k$. The values $|\alpha_0|^2$ ($\mu_0$) denote the mean photon number of the expected ideal coherent (thermal) state.}\label{fig:denmats}
\end{figure*}

The scaling of fidelity with the number of measurements for all three attenuated laser examples is plotted in Fig.~\ref{fig:COHvsTHERM}(a), with the final results (for $K=7998$) in Fig.~\ref{fig:denmats}(a), along with the complete raw datasets and parameters for the expected ideal states as insets.  Color denotes phase, either for the LO setting $\theta_k$ or for the estimated density matrix element $\rho_{mn}$. Nonzero phase values for the $\rho_{mn}$ elements are a consequence of the measured state's angular position in phase space with respect to the LO; although fluctuating randomly due to environmental perturbations, it effectively remains fixed within the 0.8~ms duration of data collection for each measurement set. Final fidelities approach unity in all cases: $F=0.958\pm 0.003$ ($\alpha_0 = 1.14-0.45\imag$, $|\alpha_0|^2=1.51$), $F=0.956\pm 0.004$ ($\alpha_0 = 0.24-1.76\imag$, $|\alpha_0|^2=3.14$), and $F=0.958\pm 0.004$ ($\alpha_0 = -2.78-0.54\imag$, $|\alpha_0|^2=7.97$).

The results for amplified spontaneous emission appear in Fig.~\ref{fig:COHvsTHERM}(b) and Fig.~\ref{fig:denmats}(b). The diagonal elements of the final estimates mirror the exponential decay of ideal thermal states, although the appreciable off-diagonal entries reveal deviations that are quantified by reduced fidelities compared to the coherent state tests: $F=0.90\pm 0.03$ ($\mu_0 = 0.68$), $F=0.82\pm 0.04$ ($\mu_0 = 1.49$), and $F=0.77\pm 0.03$ ($\mu_0 = 3.03$). While we expect the thermal states to reach comparably high fidelities with additional measurements, we have noticed a similar reduction in thermal state tomographic efficiency in simulation (see Appendix B).

\subsection{Comparing homodyne and heterodyne} \label{CHOHT}
To test the ability of our method for inference on both homodyne and heterodyne measurement schemes, as well as highlight the impact of $\eta$ selection, we can obtain homodyne data from the setup in Fig.~\ref{fig:expsetup} (for a 6~mW homodyne LO) by considering only the $x$-quadrature results; the traced-out $p$ measurement can then just be viewed as an additional 3~dB loss on the state to be characterized. Therefore, heterodyne inference with $\eta=1$ and homodyne inference with $\eta=0.5$ should match the same ground truth in our setup. Nonetheless, since this arrangement penalizes homodyne with additional loss, we also obtain a third dataset in which we double both the LO power (from 6~mW to 12~mW) and the signal power to counteract the internal beamsplitter loss in the $90^{\circ}$ optical hybrid. For coherent and thermal states, where loss only reduces average photon number~\cite{Qi:20}, the traced-out homodyne data gathered under this ``double-power'' condition reflects a state of identical statistics and average photon number to the other two cases, by taking $\eta=1$ in the likelihood. And because the vacuum noise far exceeds the electronics floor at both LO powers (see Appendix A
 )---so that detector noise can be neglected---we can directly compare tomography for three separate conditions in which the states to be inferred are matched, thus minimizing the impact of any state-dependent effects on the results. (For squeezed states or non-Gaussian states affected by loss in nontrivial ways, we acknowledge this comparison would not be valid.)

\begin{figure}
\centerline{\includegraphics[width=3.375in]{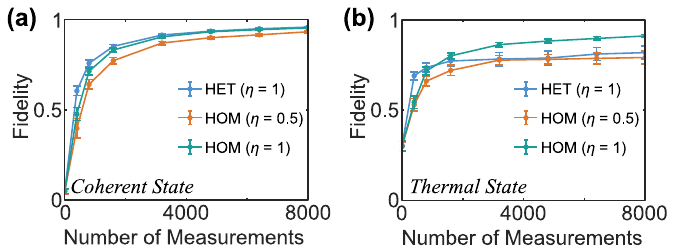}}
\caption{Comparing heterodyne (HET) and homodyne (HOM) tomography at multiple efficiencies $\eta$. (a)~Attenuated laser results. (b)~Amplified spontaneous emission. All examples have a mean photon number of $\sim$1.5. 
Homodyne datasets are obtained by considering only the $x$ quadrature data in Fig.~\ref{fig:expsetup}, with signal and LO power either equal to the heterodyne case (for HOM, $\eta=0.5$) or double (for HOM, $\eta=1$).}\label{fig:hethomcomp}
\end{figure}

Results for both an attenuated laser and amplified spontaneous emission appear in Fig.~\ref{fig:hethomcomp}, where the average photon number in all cases is $\sim$1.5. A thinning of $T=2^{14}$ and cutoff of $n_c=20$ are again used. Unsurprisingly, the lowest fidelities are obtained by the homodyne case with $\eta=0.5$: compared to heterodyne, its data are a subset that ignores the information from the entire $p$ quadrature; compared to the homodyne case with $\eta=1$, the amount of quadrature information is the same, but embedded within a higher proportion of vacuum noise. Yet the two cases with $\eta=1$ prove more subtle, showing nearly identical efficiency on the coherent state example, but an appreciable edge in favor of homodyne for the thermal state. The relative merits of homodyne and heterodyne have been explored in several contexts, including tomography of Gaussian states~\cite{Muller2016,rehacek_surmounting_2015} and the channel capacities of Gaussian receivers~\cite{Takeoka2014}. Yet although our examples here are also Gaussian, Gaussianity is not assumed in the Bayesian prior, which considers all physically allowed states. Therefore it is unclear how previous findings which assume Gaussianity \emph{a priori} would apply to our tests. In any regard, Fig.~\ref{fig:hethomcomp} serves to reveal the ability of our Bayesian approach to handle multiple measurement configurations and efficiency parameters, returning a justifiable state estimate under the prior distribution and available data---whatever the data may be.

\section{Cat state inference}
\label{sec:cat}
Our Bayesian QST workflow is specifically designed for tomography of arbitrary CV states, including non-Gaussian states of relevance to QIS. While the experimental production of such states is beyond the current capabilities of our setup in Fig.~\ref{fig:expsetup}, in this section we specifi cally test our approach on previous tomographic data from the non-Gaussian experiment described by Gerrits \emph{et al.}~\cite{Gerrits2010}, in which cat states were produced via photon subtraction of squeezed vacuum. In that work, MLE was employed on the homodyne data to recover the state produced, followed by parametric bootstrap resampling to estimate error bars. We note that Fisher information could also be used for uncertainty quantification with MLE-based QST~\cite{Rehacek2008}; in any case, the MLE and Bayesian findings are expected to agree perfectly in the asymptotic limit of infinite observations. Yet no clear relationship exists in the finite-measurement regime explored here, making it constructive to compare, so in this section the predictions of our Bayesian method to those of MLE on identical sets of quadrature samples.

\begin{figure}
\centerline{\includegraphics[width=3.25in]{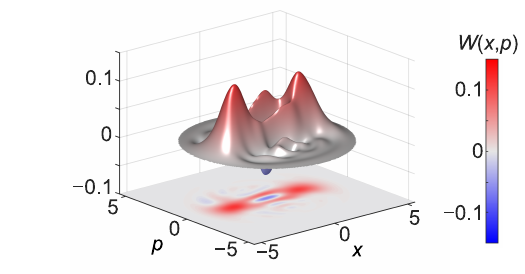}}
\caption{Wigner function of the Bayesian mean density matrix $\rho_B$ estimated from homodyne data on a three-photon-subtracted squeezed vacuum state from Ref.~\cite{Gerrits2010}. The nearest ideal cat state has $|\alpha|=1.64^{+0.09}_{-0.08}$ with fidelity $F=0.49^{+0.09}_{-0.09}$.}
\label{fig:NISTone}
\end{figure}

In Fig.~\ref{fig:NISTone}, we show the results from our Bayesian QST workflow (with $T=2^{13}$ and $n_c=10$) applied to data in Ref.~\cite{Gerrits2010} which used a transition edge sensor (TES) to implement three-photon subtraction and produce an odd cat state of the form $\ket{C_{-}(\alpha)} \propto \ket{\alpha} - \ket{-\alpha}$. To aid visual analysis, we plot the Wigner function of the Bayesian mean $\rho_B$, rather than the Fock basis elements as before; Fig.~\ref{fig:NISTone} can thus be compared directly against the MLE Wigner function in Fig.~4 of Ref.~\cite{Gerrits2010}. Adopting the same optimization approach and error bar percentiles as in Gerrits \emph{et al.}, we calculate that the nearest ideal cat state has $|\alpha|=1.64^{+0.09}_{-0.08}$, and the fidelity between that cat state and the state reconstructed from the data is $F=0.49^{+0.09}_{-0.09}$, in good agreement with the previous MLE findings $|\alpha|=1.76^{+0.02}_{-0.19}$ and $F=0.59^{+0.04}_{-0.14}$. The relatively large error bars result from the limited number of quadrature measurement values ($K=1087$), caused by the low probability of subtracting three photons. Bayesian-estimated Wigner functions for all other datasets from Ref.~\cite{Gerrits2010}, and a comparison of our findings with those from MLE, can be found in Appendix C.
For the purposes of the present investigation, however, the key point of our results is to highlight the successful application of Bayesian inference in reconstructing a non-Gaussian ground truth quantum state---a first for Bayesian QST of CV systems.

\section{Discussion}
In our experimental examples so far, we have observed appreciably lower fidelities for the inference of thermal states compared to coherent states (Figs.~\ref{fig:COHvsTHERM}--\ref{fig:hethomcomp}), which is corroborated in simulation as well [see Fig.~\ref{fig:Fsim}(b) in Appendix~B]. 
We hypothesize that this effect derives from the mixedness of thermal states: mixed states require additional parameters for their specification compared to pure states, suggesting in general the need for more measurements to reach a given error level. Thus, Bayesian inference should attain higher fidelity with respect to a ground truth pure state than to a thermal one at the same number of measurements, although further investigation into this effect would be useful.

Moving forward, as with any MCMC method, computational efficiency represents the strongest impediment toward the continued scaling to higher dimensions and multiple quantum modes. For reference, the individual chains behind the results in Figs.~\ref{fig:COHvsTHERM}--\ref{fig:NISTone} required up to a maximum of $\sim$40~hours to complete on our desktop computer (for $n_c=20$, $K=7998$, and $RT=2^{24}$ total steps)---a number which will increase both with photon cutoff $n_c$ and number of measurements $K$. Speedup through parallelization is generally hampered by the intrinsically serial nature of Markov chains, but recent developments in parallel approaches~\cite{vanderwerken_parallel_2013,calderhead_general_2014,Jacob2020, Heng2021} may present a path to estimation of vastly more complex quantum systems than hitherto possible in Bayesian QST.

Unexpectedly, the development and testing of our new method brought us into contact with several fascinating general problems in CV quantum state characterization that extend well beyond the Bayesian focus here. The fact that the relative tomographic efficiency of heterodyne and homodyne measurement varies with the ground truth signal (cf. Fig.~\ref{fig:hethomcomp}) aligns roughly with previous observations on Gaussian states~\cite{Takeoka2014, Muller2016}. Yet no systematic study comparing the efficiency of homodyne and heterodyne measurements has been performed for \emph{arbitrary} CV states. From a theoretical perspective, both homodyne and heterodyne detection are informationally complete, in that they measure the state in complete representations---namely, the Wigner~\cite{Lvovsky2009} and Husimi-$Q$~\cite{Leonhardt1993} functions, respectively. But the number of measurements required for a given level of accuracy need not be the same. In the future, therefore, it would be interesting to formally investigate the efficiency of homodyne and heterodyne detection on arbitrary quantum states, to elucidate what features of some ground truth state $\rho_g$ determine which method should be selected.

Another open question in quantum theory that surfaced in our analysis centers on distributions of random density matrices for CV quantum states. The Bures prior we have selected enjoys several commending theoretical properties, as the lone monotone metric which is both Fisher and Fubini--Study adjusted---i.e., it reduces to desired distributions in the classical and pure-state limits~\cite{Sommers2003}. However, as a DV-based distribution, Bures depends on the specific Hilbert space dimension which, in the CV tomographic case, is set somewhat arbitrarily by the photon cutoff $n_c$. Accordingly, it would prove theoretically satisfying to develop a fiducial CV density matrix distribution with a less abrupt stepwise $n_c$ dependence. This objective would seem to bear profitable overlap with other advanced CV tomographic techniques, such as the neural network approach of Ref.~\cite{Tiunov2020}; there the authors utilized a restricted Boltzmann machine ansatz, acknowledging the open question of whether practically interesting quantum states may be eliminated by this construction. Thus, we suspect that a more detailed understanding of random CV quantum state distributions could provide valuable insight into this problem as well. 

Nevertheless, we emphasize that the existence of such interesting open problems in no way diminishes the immediate practical value of Bayesian inference for CV QST. Indeed, one of the most attractive features of Bayes' theorem is its ability to incorporate any assumptions an experimenter wishes to bring to a problem, via an explicit prior distribution, and then determine the impact of both these assumptions and subsequent observations for estimation of any quantity of interest. Consequently, there is no need for a prior to align with some archetypal distributions; instead, it need only reflect whatever prior knowledge is imposed by the particular experimenter. And regardless of how this prior emerges, the Bayesian paradigm ensures that all underlying assumptions are enumerated and exposed for discussion and analysis---a situation of immense value as the field of QST continues to probe new regimes of state complexity.

\section{Conclusion}
In this work, we have introduced a workflow for Bayesian QST of generic CV states. This general method is capable of handling homodyne and heterodyne measurements and uses advanced MCMC algorithms for full Bayesian tomography with state-of-the-art computational efficiency. We experimentally prepared multiple types of Gaussian states, showing the inferred density matrices obtained from Bayesian estimation. Moreover, we used legacy measurement data of cat states to verify our workflow's ability to reconstruct non-Gaussian  states as well. This powerful computational analysis tool could be helpful in the precise characterization of up-and-coming CV quantum computing hardware~\cite{larsen2021deterministic}, CV quantum networking~\cite{guo2020distributed}, and CV quantum sensors~\cite{QSRev}, where it is crucial to accurately characterize the states produced.

\section*{Appendix A: Experimental Procedures} \label{sec:SMcal}
\begin{figure}
\centerline{\includegraphics[scale=0.9]{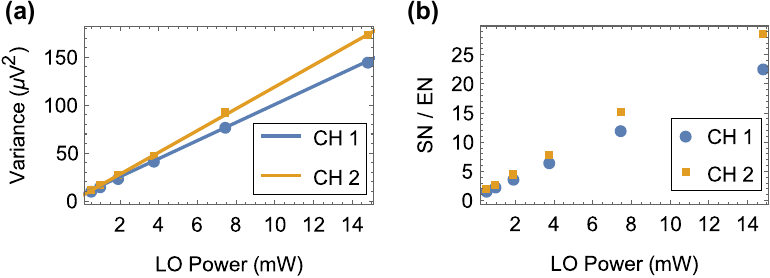}}
\caption{Shot-noise calibration. Channel 1 data (fit) correspond to the blue circle (line). Similarly, channel 2 data (fit) correspond to the orange square (line). (a)~Shot-noise variance versus LO power. From the fit: R$^2_\text{CH1}$=0.9998 and R$^2_\text{CH2}$=0.9997. (b)~Ratio of shot-noise variance to electronics noise variance versus LO power. We normally operate at about 12 mW to balance linearity and electronics noise.}\label{fig:SNcal}
\end{figure}

For polarization alignment, we found it necessary to minimize the cross-talk between the fast- and slow-axes of the polarization-maintaining-fiber at the couplings before and after the phase modulator, using polarizers and polarization controllers
; if the LO polarization is not aligned perfectly to the axis of the phase modulator, it is possible for the phase modulator to rotate the polarization of the LO, which undesirably makes the shot noise oscillate with time. To calibrate the shot noise, for a given LO power, we record two million samples (spanning 0.8 ms at our 2.5~GHz sampling rate) for each output of the heterodyne detector. At a spacing of 250 samples (100 ns) of each output channel, adjusted by the known electronic delay mismatch (separately calibrated with bright pulses), we average groups of four adjacent samples and save the average values as single quadrature points. We perform this procedure for the whole data file, which produces 7998 measurements, from which we calculate the mean, $SN_{\text{m}}$, and shot-noise variance, $SN_{\text{var}}$, for each detector channel. We calibrate the shot noise for a range of relevant optical powers [see Fig.~\ref{fig:SNcal}(a)]. We find a linear operating range from 1--15 mW, as measured at the input to the heterodyne detector, and primarily operate at $\sim$12 mW to be far from the electronics noise. We also measured the electronics noise and plot the ratio of the shot-noise variance and the electronics noise variance in Fig.~\ref{fig:SNcal}(b), which at our primary operating point is about 17.

To calibrate the phase modulator's applied voltage for a total phase swing of $2\pi$, we send in a bright coherent input signal and adjust the amplitude of a 5~kHz sawtooth voltage signal to achieve a continuous sinewave output on the oscilloscope. Any peak-to-peak amplitude not corresponding to a multiple of $2\pi$ will produce a discontinuous jump at the falling edge of each sawtooth, and when exactly $2\pi$ is achieved, a single sinewave period will be traversed during a single ramp. A square-wave synchronization signal from the function generator is also sent to the oscilloscope and saved to provide a marker for the beginning and end of each ramp. Assuming linear modulator operation, we then know the phase change between each sample.

When there is an input signal, we convert our collected data from voltage into dimensionless quadrature units, using the shot-noise calibration. As before, we average adjacent groups of four oscilloscope samples, spaced every 250 samples, 
to provide individual voltage points $V_{\text{avg}}[k]$. For our total 0.8~ms record length, we therefore complete a full $2\pi$ phase sweep four times. We then normalize these values such that a vacuum input has mean zero ($\braket{x}_\text{vac}=\braket{p}_\text{vac}=0$) and variance 1/2 ($\Delta x^2_\text{vac} = \Delta p^2_\text{vac} = 1/2$). Thus a single pair of quadratures $(x_k,p_k)$ is given by 
\begin{equation}
    x_k = (V_{X,\text{avg}}[k]-SN_{\text{X,m}})\sqrt{\frac{1/2}{SN_{\text{X,var}}}}
\end{equation}
\begin{equation}
    p_k = (V_{P,\text{avg}}[k]-SN_{\text{P,m}})\sqrt{\frac{1/2}{SN_{\text{P,var}}}}
\end{equation}
for all $k\in\{1,2,...,K\}$. These normalized points can then be input into our tomographic method.

\section*{Appendix B: Simulations} \label{sec:SMsim}
For a further test on our workflow, we produced simulated data for several representative classes of states and perform Bayesian inference as a function of measurement number. As ground truth $\rho_g$ ($\rho_g=\ket{\psi_g}\bra{\psi_g}$ if pure), we consider coherent states 
\begin{equation}
\label{eq:coh}
\ket{\psi_g} = e^{-|\alpha_0|^2/2} \sum_n \frac{\alpha_0^n}{\sqrt{n!}}\ket{n},
\end{equation}
thermal states
\begin{equation}\rho_g = \sum_n \frac{\mu^n}{(1+\mu)^{n+1}} \ket{n}\bra{n},
\end{equation} 
squeezed vacuum 
\begin{equation}\ket{\psi_g} = \frac{1}{\sqrt{\cosh r}} \sum_n \frac{\sqrt{(2n)!}}{n!} \left(\frac{1}{2}\tanh r \right)^n \ket{2n},
\end{equation}
and Fock states 
\begin{equation}
\ket{\psi_g}=\ket{n_0}.
\end{equation}
Selecting an LO phase $\theta_k$ uniformly at random from a $2\pi$ interval, and taking $\eta=1$ for these tests, we computed the corresponding probability density [Eq.~(\ref{eq:LLhom}) or (\ref{eq:LLhet})] 
over a predefined grid of points (we chose a resolution $\Delta x = \Delta p = 0.07$) and then drew a single result $x_k$ or $(x_k,p_k)$ according to a multinomial distribution, repeating this procedure $K=8000$ times.

We simulated both homodyne and heterodyne tomography for four ground truth states from each class, with parameters chosen for a mean photon number $\braket{n}\in\{0,1,2,3\}$. With our chosen cutoff $n_c=10$, the truncation error $\epsilon = 1-\sum_{n=0}^{n_c} \braket{n|\rho_g|n}$ is less than 0.07 for all cases; however, since the same $n_c$ is used in both data generation and Bayesian estimation, the only impact of this error is to modify the ground truth state in a known way, and thus not the inference accuracy. For each state and measurement subset of size $K\in\{1,400,800,1600,3200,6400,8000\}$, we perform MCMC with a thinning of $T=2^{11}$ and obtain a set of samples $\{\rho^{(r)}\}$ from which we compute the mean and standard deviation of sample fidelity 
\begin{equation}
    F^{(r)}=\left(\Tr \sqrt{\sqrt{\rho_g} \rho^{(r)} \sqrt{\rho_g}} \right)^2
\end{equation}
for plotting.

\begin{figure*}[tb!]
\centerline{\includegraphics[width=5.25in]{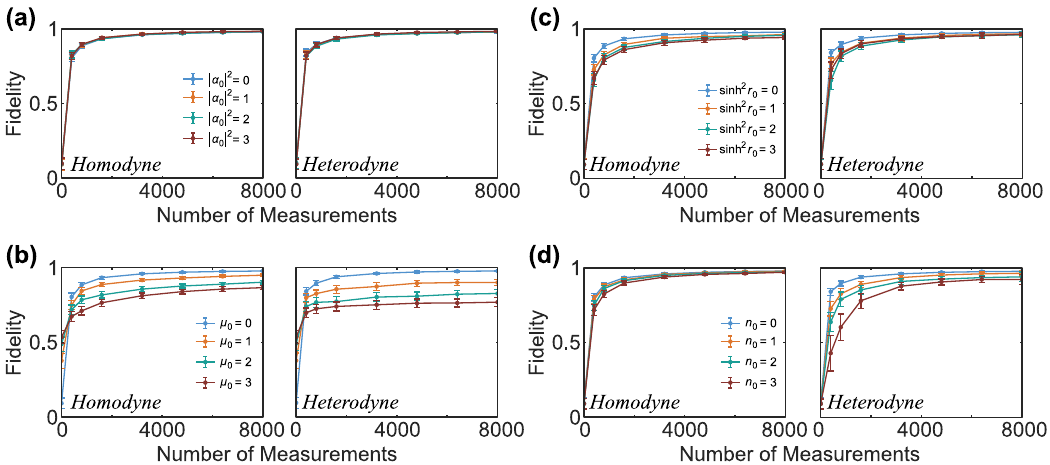}}
\caption{Inference results from simulated datasets. (a)~Coherent states. (b)~Thermal states. (c)~Squeezed vacuum. (d)~Fock states. A photon number cutoff of $n_c=10$ is utilized for both simulation and inference.}\label{fig:Fsim}
\end{figure*}

In Fig.~\ref{fig:Fsim}, we show the progression of the fidelity as more measurements---i.e., repeated preparations of the input $\rho_g$---are included in the likelihood. In all cases, the fidelity increases monotonically with $K$ as expected, although the rate varies with both ground truth state and measurement type. The thermal states show the slowest convergence of all state families, consistent with experimental findings in Figs.~\ref{fig:COHvsTHERM} and \ref{fig:hethomcomp}.
The Fock states (d) experience the most striking difference between homodyne and heterodyne measurement schemes, with the former converging in fidelity significantly more rapidly; the other states (a--c) appear less sensitive to the chosen measurement type
, although the thermal states still evince noticeably lower tomographic efficiency for heterodyne.

\section*{Appendix C: Complete Cat State Results} \label{sec:SMcat}
\begin{figure*}[b!]
\centerline{\includegraphics[width=5.25in]{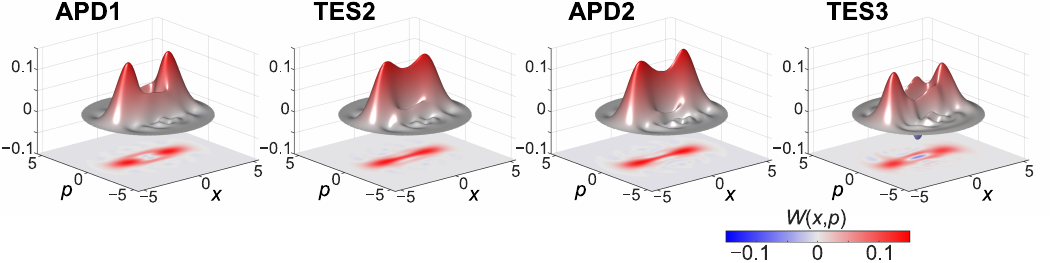}}
\caption{Wigner functions of Bayesian mean density matrices for all four experimental datasets from Ref.~\cite{Gerrits2010}. Labels denote the type of detector used (APD or TES) and the number of photons subtracted (1--3).}\label{fig:NISTall}
\end{figure*}

In Fig.~\ref{fig:NISTall}, we show the results from our Bayesian QST workflow for all four datasets in Ref.~\cite{Gerrits2010}, which used avalanche photodiodes (APDs) or a TES to implement one-, two-, and three-photon subtraction on squeezed vacuum to herald coherent-state superpositions (cat states), according to the ideal form
\begin{equation}
\label{eq:cat}
\ket{C_\pm (\alpha)} = \frac{1}{\sqrt{2(1\pm e^{-2|\alpha|^2})}}\left(\ket{\alpha} \pm \ket{-\alpha} \right),
\end{equation}
where $\ket{\pm\alpha}$ are as defined in Eq.~(\ref{eq:coh}), and $\pm$ corresponds to the even (odd) cat state. The Wigner function for a given density matrix $\rho$ is defined under our $\hbar=1$ convention as
\begin{equation}
\label{eq:Wigner}
W(x,p) = \sum_m\sum_n \rho_{mn} \frac{(-1)^n}{\pi} \sqrt{\frac{2^m n!}{2^n m!}} e^{-(x^2+p^2)}[-(x-\imag p)]^{m-n} L_n^{m-n}(2(x^2 + p^2)),
\end{equation}
where $L_n^{m-n}(\cdot)$ denotes the generalized Laguerre function.

Following the procedure of Gerrits \emph{et al.}, to each MCMC sample $\rho^{(r)}$ we assign a cat state amplitude $\alpha^{(r)}$ as that with maximum overlap:
\begin{equation}
\label{eq:argmax}
\alpha^{(r)} =   \underset{\alpha}{\arg\max} \braket{C_\pm(\alpha)|\rho^{(r)}|C_\pm(\alpha)},
\end{equation}
so that the sample fidelity follows as
\begin{equation}
\label{eq:sampF}
F^{(r)} = \braket{C_\pm(\alpha^{(r)})|\rho^{(r)}|C_\pm(\alpha^{(r)})}.
\end{equation}
We then report $|\alpha|$ and $F$ estimates as $B^{U-B}_{L-B}$, where $B$ is the Bayesian mean, $U$ the 84th percentile, and $L$ the 16th percentile, again following Ref.~\cite{Gerrits2010}.
Wigner functions of the Bayesian mean density matrices $\rho_B$ for all cases appear in Fig.~\ref{fig:NISTall}. Our procedure assumes a system efficiency $\eta=0.853$ and cutoff $n_c=10$, for which the truncation error for all expected cat states is less than $7\times 10^{-6}$; $R=1024$ samples are again used for calculations, with a thinning value of $T=2^{13}$ for all cases except APD1, which required $T=2^{15}$ (due to the larger number of quadrature samples $K$).

\begin{figure}
\centerline{\includegraphics[width=5.25in]{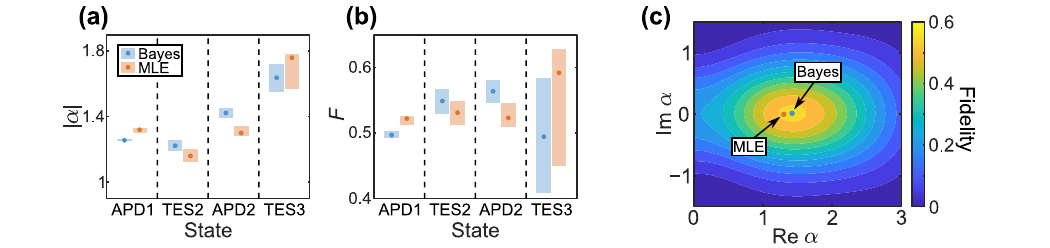}}
\caption{Comparison of analysis techniques on experimental cat state data. (a)~Estimated cat state amplitude $|\alpha|$. (b)~Estimated fidelity $F$. The dots represent either the Bayesian mean calculated here or the MLE point estimate from Ref.~\cite{Gerrits2010}, and the boxes enclose the 16th--84th percentiles. (c)~Fidelity between an even cat state $\ket{C_+(\alpha)}$ and the Bayesian mean $\rho_B$ for the APD2 dataset. Points show the corresponding estimates from (a).}\label{fig:compNIST}
\end{figure}

Estimates for $|\alpha|$ and $F$ with uncertainties are plotted in Fig.~\ref{fig:compNIST}, along with the MLE findings of Gerrits \emph{et al.} for comparison. The variation in error box length follows from the diversity in dataset size. APD1 has 324 510 quadrature samples, TES2 has 24 790, APD2 has 41 223, and TES3 has 1087. Both methods show very similar uncertainties for each state, highlighting the congruity between Bayesian uncertainty quantification and parametric bootstrap resampling in these examples. Agreement in the absolute values is also generally good, though the estimated ranges do not overlap in all cases. Given the use of experimental data and differences between Bayesian inference and MLE, there is no fundamental reason to expect perfect agreement. Nevertheless, we do note that the $\alpha$ landscape for optimization via Eq.~(\ref{eq:argmax}) is quite flat. For example, in Fig.~\ref{fig:compNIST}(c) we show the fidelity between an ideal even cat state of variable complex amplitude $\alpha$ and the Bayesian mean ($F= \braket{C_+(\alpha)|\rho_B|C_+(\alpha)}$) for the APD2 result $\rho_B$. Despite the appreciable separation between the Bayesian and MLE estimates for $|\alpha|$, both cat state estimates (shown as dots in the contour plot) possess very similar overlap with $\rho_B$ ($F=0.56$ when rounded), indicating closer similarity between the Bayesian and MLE estimates than the uncertainty intervals in Fig.~\ref{fig:compNIST}(a) might suggest. This points to the general difficulty in assigning a unique cat state to a particular density matrix $\rho^{(r)}$.

\begin{backmatter}

\section*{Funding}
U.S. Department of Energy, Office of Science, Advanced  Scientific Computing Research, Transparent Optical Quantum Networks
for Distributed Science Program and Early Career Research Program (Field Work Proposals ERKJ355 and ERKJ353).

\section*{Acknowledgments}
 We are grateful to T. Gerrits for providing the data from Ref.~\cite{Gerrits2010} for use in our analyses. We thank B.~T. Kirby, S. Guha, C.~N. Gagatsos, and A.~J. Pizzimenti for valuable discussions. This work was performed at Oak Ridge National Laboratory, operated by UT-Battelle for the U.S. Department of Energy under contract no. DE-AC05-00OR22725.

\section*{Disclosures}
The authors declare no conflicts of interest.

\section*{Data Availability Statement}
Data underlying the results presented in this paper are not publicly available at this time but may be obtained from the authors upon reasonable request.

\end{backmatter}



\end{document}